\documentclass{SCGE}
\usepackage[colorlinks,linkcolor=blue,citecolor=blue,urlcolor=blue]{hyperref}

\let\citedash\relax
\makeatletter \providecommand{\citedash}{\hbox{-}\penalty\@m}
\makeatother

\begin{document}
\begin{picture}(0,0){\rm
\put(0,-20){\makebox[160truemm][l]{\bf {\sanhao\raisebox{2pt}{.}}
Article  {\sanhao\raisebox{1.5pt}{.}}}}}
\put(0,-34){\jiuwuhao {\textcolor[rgb]{0.5,0.5,0.5}{\sf 
}}}
\end{picture}

\def\bm{\boldsymbol}implementation

\def\dl{\displaystyle}
\def\du{\end{document}}
\def\d{{\rm d}}
\def\e{{\rm e}}
\def\i{{\rm i}}

\Year{2017} %
\Month{ } %
\Vol{61} 
\No{1} 
\BeginPage{1} 
\AuthorMark{{\rm Qing Lin}}  
\DOI{} 
\ArtNo{000000}

\title[Multiple multi-control unitary operations: implementation and applications]{Multiple multi-control unitary operations: implementation and applications}

\author[*]{Qing Lin}{}

\address[{\rm}]{Fujian Key Laboratory of Light Propagation and Transformation, College of Information Science and Engineering, 
Huaqiao University,
Xiamen 361021, China}

\maketitle \vspace{-3.5mm}{\footnotesize\begin{center} Received January 1, 2016; accepted January 1, 2016; published online January 1, 2016
\end{center}}\vspace*{-5mm}

\begin{center}
\rule{16.5cm}{0.4pt}
\parbox{16.5cm}
{\begin{abstract} 
How to implement a computation task efficiently is the central problem in quantum computation science. For a quantum circuit, the multi-control unitary operations are the very important components. We present an extremely efficient approach to implement multiple multi-control unitary operations directly without any decompositions to CNOT gates and single-photon gates. The necessary two-photon operations could be reduced from $\mathcal{O}(n^3)$ with the traditional decomposition approach to $\mathcal{O}(n)$ with the present approach, which will greatly relax the requirement resources and make this approach much feasible for large scale quantum computation. Moreover, the potential application to the ($n$-$k$)-uniform hypergraph state generation is proposed.
\end{abstract}}
\end{center}\vspace*{-0.6cm}

\begin{center}
\parbox{16.5cm}
{\bf\jiuhao multi-control unitary operation, cross phase modulation, c-path gate, merging gate, hypergraph state}
\end{center}

\begin{center}
{}
\CITA    
\end{center}

\textwidth=178truemm \textheight=236truemm

\wuhao\vspace*{1.5mm}

\begin{multicols}{2}

\renewcommand{\baselinestretch}{1.08} \baselineskip 12.2pt\parindent=10.8pt

\renewcommand{\thefootnote}

\section{Introduction}\label{sec:1}
Quantum computation attracts more and more attentions during the last two decades. It can provide more powerful and extremely faster computation than the classical computation \cite{Nielsen}. After tremendous effort has been devoted for many years in quantum information science, people now turn to the topic of large scale quantum computation. In 2009, 14-qubit entanglement has been reported in trapped ions system \cite{14et}. Later in 2012, based on spontaneous parametric down-conversion, eight-photon GHZ states have been created by Huang et. al.  \cite{8photon1} and Yao et al. \cite{8photon2}, respectively. After four years, Wang et. al. reported an excited development about ten-photon entanglement \cite{10photon}. However, the probability problem is inevitable for linear optical quantum computation \cite{KLM}. The photons  themselves do not interacted with each other, therefore one have to achieve the necessary interaction for two-photon \noindent\rule{2.5cm}{0.4pt}\\[0.1mm]{\qihao *Corresponding author (email: qlin@hqu.edu.cn)}\\
\noindent operations, such as CNOT, etc., through the postselection  technique, which will result in the nondeterministical two-photon operations with only linear optical elements. For example, the success probability of a CNOT gate is only $1/4$ \cite{CNOT1,CNOT2}. With the increasing of involved photon number and operation number, the success probability will be decreased exponentially. How to implement the operations efficiently becomes a central task of optical quantum computation.

On the other hand, the cross phase modulation (XPM) technique developed recently should be a useful supplement to linear optical technology for quantum computation. With the help of XPM technique, the construction of Bell state analysis \cite{XPM1}, CNOT gate \cite{XPM2, lin1}, and even Toffoli gate \cite{lin2, lin3}, etc., could be nearly deterministic. It provides a new router to quantum computation \cite{XPM3, XPM4, XPM5, lin4, lin5}. Especially, with the element gates, called controlled-path (c-path) and merging gates proposed in Refs. \cite{lin1, lin2, lin3}, the implementation of quantum computation could be direct without any decompositions to CNOT gates and single photon gates, but highly efficient with less resources. For example, the necessary resources for the multi-control gates, such as general Toffoli gate, Fredkin gate, etc., are only linear increasing ($\mathcal{O}(n)$) with the involved photon number \cite{lin2, lin3}, compared with the polynomial increasing ($\mathcal{O}(n^2)$) in other proposals \cite{Barenco}. Even for a general unitary operations, the approach with c-path and merging gates (called CPM approach below) could exponentially relax the requirement on the resources, including the number of operations, the ancilla photons and coherent beams, etc \cite{lin5}.

The former works with the XPM technique are either focus on some specific gates, such as CNOT \cite{XPM2,lin1}, Toffoli, Fredkin gates \cite{lin2, lin3, lin4}, etc., or the general operations, such as the general unitary operations \cite{lin5}. A problem raises naturally that whether one can further reduce the complexity or not, if a quantum circuit includes some special structures. Here we consider a quantum circuit constructed by a series of multi-control operations. As mentioned above, a multi-control (($n$-$1$)-control) operation requires $\mathcal{O}(n)$ c-path and merging gates \cite{lin2, lin3}. We will show below, with the new design of c-path gate, even for multiple ($n$) multi-control operations, $\mathcal{O}(n)$ c-path and merging gates are enough. In other words, the increasing with the involved photon number is preserved to be linear. Moreover, this special operation can be applied to generate the hypergraph states \cite{hyper1,hyper2,hyper3,hyper4,hyper5}, which is a very important state for measurement based quantum computation.

The rest of the paper is organized as follows. In Sec. \ref{sec:3}, we introduce a new c-path gate, which will be used as the element gate to implement multi-photon gates, associated with the original c-path and merging gate. As an example in Sec. \ref{sec:4}, we discuss the implementation of four triple-control unitary operations, which can be generalized to multiple ($n$-$1$)-control unitary operations in Sec. \ref{sec:5}. To show the efficiency of the present approach, we compare the implementation complexity with the former approaches in Sec. \ref{sec:6}. In addition, the approach is generalized in Sec. \ref{sec:7} to implement the multiple ($n$-$k$)-control unitary operations, and the corresponding potential applications are discussed in Sec. \ref{sec:8}. The final part is for discussion and conclusion remark.

\section{The controlled-path gate and merging gate}\label{sec:3}

The original controlled-path (c-path) gate was firstly introduced in Ref. \cite{lin1}, and then developed to be more feasible in Refs. \cite{lin2, lin3, lin4, lin5}. This gate can be implemented using only linear optics as well and had been demonstrated in realistic experiment \cite{cpathe1}. Moreover, it can be used in experiments to calculate unknown eigenvalues \cite{cpathe2}, compress quantum data \cite{cpathe3}, or implement a quantum Fredkin gate \cite{cpathe4}, and widely used in theoretical schemes to generate graph states \cite{lin6, lin7}, W states \cite{lin8}, Dicke states \cite{lin9, lin10}, etc. 

The control photon of original c-path gate is encoded by its polarization information as, $\left\vert 0\right\rangle \equiv \left\vert H\right\rangle$ and $\left\vert 1\right\rangle\equiv\left\vert V\right\rangle$. However, when we discuss the implementation of multiple multi-control unitary operations in below, the control photon has some spatial modes, e.g, $8$, but we only use part of them as the control signal. That does mean the control signal sometime contains the one single photon, and sometime is a vacuum state. In other words, the control signal is encoded by the photon number, i.e., the control signal $\left\vert 0\right\rangle$ denotes a vacuum state and $\left\vert 1\right\rangle$ denotes the quantum state containing one single photon. The desired c-path gate is required to implement the following transformation,
\begin{align}
&\left\vert \Psi_1\right\rangle\left\vert 0\right\rangle^C\left\vert \Phi_1\right\rangle^T+\left\vert \Psi_2\right\rangle\left\vert 1\right\rangle^C\left\vert \Phi_2\right\rangle^T\nonumber\\
\rightarrow&\left\vert \Psi_1\right\rangle\left\vert 0\right\rangle^C\left\vert \Phi_1\right\rangle_1^T+\left\vert \Psi_2\right\rangle\left\vert 1\right\rangle^C\left\vert \Phi_2\right\rangle_2^T, \label{ncpath}
\end{align}
where the states $\left\vert \Psi_{1(2)}\right\rangle$ denotes the other components of the multi-photon state. The target single photon denoted by a superscript $T$ will be separated into one spatial mode when the single photon appears on the control spatial modes ($\left\vert 1\right\rangle^C$), while it will be separated into another one spatial mode when the single photon appears on the other spatial modes which haven't been used as control modes ($\left\vert 0\right\rangle^C$). 

\begin{figure}[H]\centering
\includegraphics[scale=0.42]{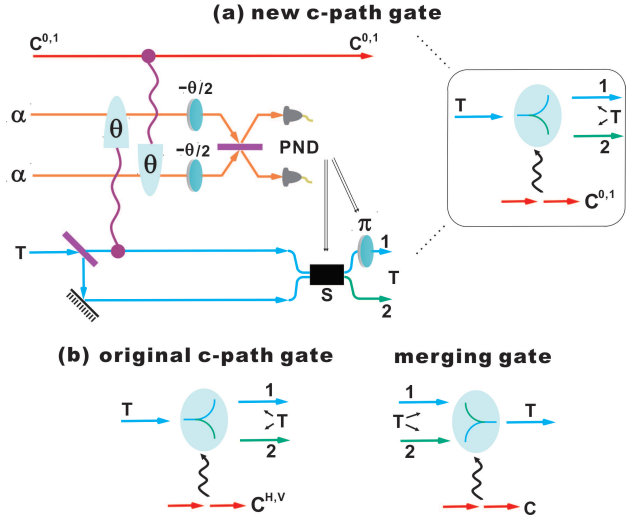}
\caption{(Color online) (a) the new c-path gate. The target single photon will be transformed to spatial mode $2$, when the control mode containing one single photon; otherwise, it will be transformed to spatial mode $1$. (b) the original c-path and merging gate. The detailed design of these two gates can be found in Appendix. For comparison, the control mode of the original c-path gate is denoted by superscripts ($H, V$), while that of the new c-path gate is denoted by superscripts ($0,1$). For details, see text.} 
\label{fig1}
\end{figure}

Clearly, the original c-path gate cannot be used to implement the above operation. We now design a new c-path gate shown in Fig. \ref{fig1}(a) to implement the above operation. Firstly, the target photon is injected into a 50:50 beam splitter (BS). After that the control and target photon are interacted with two coherent beams $\left\vert \alpha\right\rangle_{cs}\left\vert \alpha\right\rangle_{cs}$ through the XPM processes indicated in Fig. \ref{fig1}(a). The initial state will evolved to the following state,
\begin{align}
&\frac{1}{\sqrt{2}} \left(\left\vert \Psi_1\right\rangle\left\vert 0\right\rangle^C\left\vert \Phi_1\right\rangle_1^T\left\vert \alpha e^{i\theta}\right\rangle_{cs}\left\vert \alpha\right\rangle_{cs}+\left\vert \Psi_1\right\rangle\left\vert 0\right\rangle^C\left\vert \Phi_1\right\rangle_2^T\left\vert \alpha\right\rangle_{cs}\left\vert \alpha\right\rangle_{cs}\right. \nonumber\\
+&\left.\left\vert \Psi_2\right\rangle\left\vert 1\right\rangle^C\left\vert \Phi_2\right\rangle_1^T\left\vert \alpha e^{i\theta}\right\rangle_{cs}\left\vert \alpha e^{i\theta}\right\rangle_{cs}+\left\vert \Psi_2\right\rangle\left\vert 1\right\rangle^C\left\vert \Phi_2\right\rangle_2^T\left\vert \alpha \right\rangle_{cs}\left\vert \alpha e^{i\theta}\right\rangle_{cs}\right),
\end{align}
where the subscripts $1, 2$ outside the bracket denote the spatial modes of the target photon respectively. Two phase shifts $-\theta/2$ are applied to the two coherent beams respectively, yielding the following state,
\begin{align}
&\frac{1}{\sqrt{2}} \left(\left\vert \Psi_1\right\rangle\left\vert 0\right\rangle^C\left\vert \Phi_1\right\rangle_1^T\left\vert \alpha e^{i\theta/2}\right\rangle_{cs}\left\vert \alpha e^{-i\theta/2}\right\rangle_{cs}+\left\vert \Psi_1\right\rangle\left\vert 0\right\rangle^C\left\vert \Phi_1\right\rangle_2^T \right.\nonumber\\
&\otimes\left\vert \alpha e^{-i\theta/2}\right\rangle_{cs}\left\vert \alpha e^{-i\theta/2}\right\rangle_{cs}+\left\vert \Psi_2\right\rangle\left\vert 1\right\rangle^C\left\vert \Phi_2\right\rangle_1^T\left\vert \alpha e^{i\theta/2}\right\rangle_{cs}\left\vert \alpha e^{i\theta/2}\right\rangle_{cs} \nonumber\\
&\left.+\left\vert \Psi_2\right\rangle\left\vert 1\right\rangle^C\left\vert \Phi_2\right\rangle_2^T\left\vert \alpha e^{-i\theta/2}\right\rangle_{cs}\left\vert \alpha e^{i\theta/2}\right\rangle_{cs}\right).
\end{align}
Let the two coherent beams to be interfered on a 50:50 BS, and one can get the following state,
\begin{align}
&\frac{1}{\sqrt{2}} \left(\left\vert \Psi_1\right\rangle\left\vert 0\right\rangle^C\left\vert \Phi_1\right\rangle_1^T\left\vert \alpha^+\right\rangle_{cs}\left\vert \alpha^-\right\rangle_{cs}+\left\vert \Psi_1\right\rangle\left\vert 0\right\rangle^C\left\vert \Phi_1\right\rangle_2^T \right. \nonumber\\
&\otimes\left\vert \sqrt{2}\alpha e^{-i\theta/2}\right\rangle_{cs}\left\vert 0\right\rangle_{cs}+\left\vert \Psi_2\right\rangle\left\vert 1\right\rangle^C\left\vert \Phi_2\right\rangle_1^T\left\vert \sqrt{2}\alpha e^{i\theta/2}\right\rangle_{cs}\left\vert 0\right\rangle_{cs} \nonumber\\
&\left.+\left\vert \Psi_2\right\rangle\left\vert 1\right\rangle^C\left\vert \Phi_2\right\rangle_2^T\left\vert \alpha^+\right\rangle_{cs}\left\vert -\alpha^-\right\rangle_{cs}\right). \label{sbd}
\end{align}
where $\left\vert \alpha^-\right\rangle=\left\vert i\sqrt{2}\alpha \sin(\theta/2)\right\rangle$ and $\left\vert \alpha^+\right\rangle=\left\vert \sqrt{2}\alpha \cos(\theta/2)\right\rangle$. 

Now, we detect the second coherent beam with a photon number resolving detector (PND). The PND can be implemented by the XPM technique as well, which had been proposed in former works \cite{lin2,lin3}. If $k$ single photons are registered, one can get the following state,
\begin{align}
e^{ik\pi/2}\left\vert \Psi_1\right\rangle\left\vert 0\right\rangle^C\left\vert \Phi_1\right\rangle_1^T\left\vert \alpha^+\right\rangle_{cs}+e^{-ik\pi/2}\left\vert \Psi_2\right\rangle\left\vert 1\right\rangle^C\left\vert \Phi_2\right\rangle_2^T\left\vert \alpha^+\right\rangle_{cs}
\end{align} 
 In this case, the first coherent state in above two components is the same, and $\sqrt{2}\alpha \cos(\theta/2)\sim \sqrt{2}\alpha$ due to $\theta \ll 1$, e.g., $\theta \sim 10^{-5}$ for weak Kerr nonlinearity, so it can be recycled. The unwanted phase shifts $e^{ik\pi/2}$ and $e^{-ik\pi/2}$ can be removed by the additional phase shifts applied to the first spatial mode controlled by the detection $k$ of the PND through the classical feedforward. After that, the following desired state can be achieved,
\begin{align}
\left\vert \Psi_1\right\rangle\left\vert 0\right\rangle^C\left\vert \Phi_1\right\rangle_1^T+\left\vert \Psi_2\right\rangle\left\vert 1\right\rangle^C\left\vert \Phi_2\right\rangle_2^T, \label{desire}
\end{align} 
 i.e., the target photon will be transformed to the first spatial mode $1$ when the controlled state is a vacuum state; otherwise, the target photon will be transformed to the second spatial mode $2$.
 
On the other hand, if none single photons is registered by the PND, the state in Eq. (\ref{sbd}) will be collapsed to the following state,
\begin{align}
\left\vert \Psi_1\right\rangle\left\vert 0\right\rangle^C\left\vert \Phi_1\right\rangle_2^T\left\vert \sqrt{2}\alpha e^{-i\theta/2}\right\rangle_{cs}+\left\vert \Psi_2\right\rangle\left\vert 1\right\rangle^C\left\vert \Phi_2\right\rangle_1^T\left\vert \sqrt{2}\alpha e^{i\theta/2}\right\rangle_{cs}.
\end{align}
Since the rest first coherent beam containing opposite phase shifts, then it should be removed further. We detect the first coherent beam with an additional PND. If $m$ single photons are registered, one can obtain the following state,
\begin{align}
e^{-im\theta/2}\left\vert \Psi_1\right\rangle\left\vert 0\right\rangle^C\left\vert \Phi_1\right\rangle_2^T+e^{im\theta/2}\left\vert \Psi_2\right\rangle\left\vert 1\right\rangle^C\left\vert \Phi_2\right\rangle_1^T.
\end{align}
The unwanted phase shifts $e^{-im\theta/2}$ and $e^{im\theta/2}$ can be removed similarly and the above state can be transformed to the desired state in Eq. (\ref{desire}) by the switch controlled by the detection of first PND through the classical feedforward as well.

Since the original c-path and merging gate will be used in below as well, then we describe what it can work without details here and the design of the c-path and merging gate can be found in Appendix. The original c-path gate is to implement the following transformation, 
\begin{align}
&\left\vert \Psi_1\right\rangle\left\vert H\right\rangle^C \left\vert \phi\right\rangle^T+\left\vert \Psi_2\right\rangle\left\vert V\right\rangle^C \left\vert \psi\right\rangle^T\nonumber\\
\rightarrow&\left\vert \Psi_1\right\rangle\left\vert H\right\rangle^C \left\vert \phi\right\rangle^T_1+\left\vert \Psi_2\right\rangle\left\vert V\right\rangle^C \left\vert \psi\right\rangle^T_2, \label{cpe}
\end{align} 
i.e., the target photon will be separated into two spatial modes $1, 2$, which depend on the polarizations of the control single photon, but not the photon number in the new c-path gate shown in Eq. (\ref{ncpath}). Moreover, for a complete framework of quantum computation, we need the inverse transformation of the original or new c-path gate. As shown in Appendix, the necessary inverse transformation can be completed by the same merging gate \cite{lin2,lin3,lin4,lin5}, which can merge the two spatial modes of the target single photon back to one without changing anything else, i.e., implementing the inverse transformation of Eq. (\ref{ncpath}) or Eq. (\ref{cpe}). Furthermore, we should note here that, in the c-path gate (including original and new c-path gate) or merging gate, the spatial modes of the target single photon could be more than $2$. The c-path gate will separate the $n$ spatial modes of target photon into $2n$ spatial modes, while the merging gate will merge the $2n$ spatial modes back to $n$ spatial modes. In one word, the c-path and merging gate can work well in the case of more than $2$ spatial modes, which provides a flexible way to use them.

\section{implementation of four triple-control unitary operations}\label{sec:4}
\begin{figure*}[t]\centering
\includegraphics[scale=0.5]{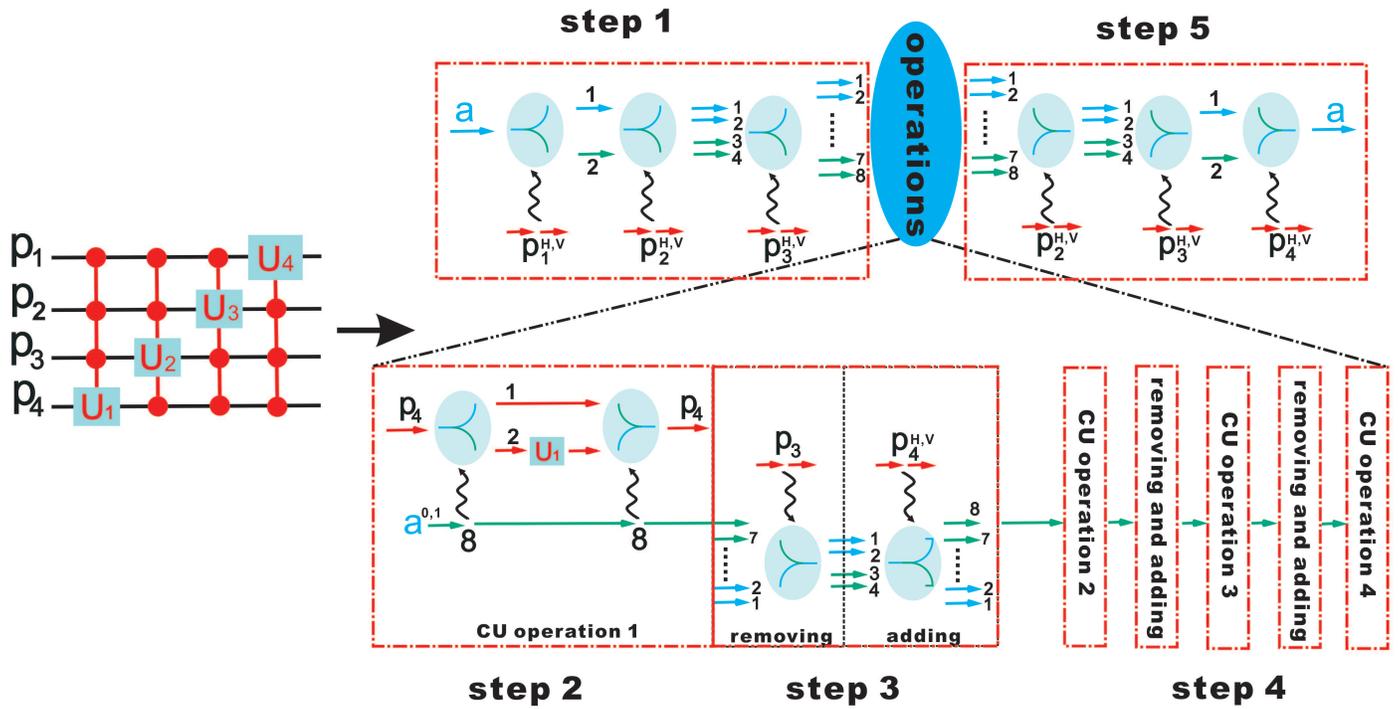}
\caption{(Color online) The implementation of four triple-control unitary operations. The processes are separated into 5 steps: (1). an ancilla single photon is introduced, which will be separated into $8$ spatial modes controlled by the first three single photons $P_1$, $P_2$, $P_3$ through three original c-path gates respectively; (2). using the spatial mode $8$ of the ancilla single photon to control the first target single photon $P_4$ by the new c-path gate, the first control unitary operation will be implemented, associated with the single photon unitary operation $U_1$ applied to the spatial mode $2$ of target single photon and the merging gate; (3). to implement the second controlled unitary operation, the information of single photon $P_3$ should be removed from the ancilla single photon (by a merging gate), while the information of single photon $P_4$ should be added to the ancilla single photon (by an original c-path gate); (4). following the similar processes (operation 2, 3, 4), the desired four triple-control unitary operations can be implemented; (5). the ancilla single photon can be disentangled from the four-photon state by the three merging gates.} 
\label{fig2}
\end{figure*}
With the new c-path gates and the original c-path and merging gates, the implementation of multiple multi-control unitary operations can be more efficiently than the one using only original c-path and merging gates. The corresponding discussion of efficiency can be found in Sec. \ref{sec:6}. Without loss generality, we use the implementation of four triple-control unitary operations displayed in Fig. \ref{fig2} as an example and this approach can be generalized easily. The general four-photon state can be described as follows,
\begin{align}
&A_1\left\vert HHHH\right\rangle+A_2\left\vert HHHV\right\rangle+A_3\left\vert HHVH\right\rangle+A_4\left\vert HHVV\right\rangle\nonumber\\
+&A_5\left\vert HVHH\right\rangle+A_6\left\vert HVHV\right\rangle+A_7\left\vert HVVH\right\rangle+A_8\left\vert HVVV\right\rangle\nonumber\\
+&A_9\left\vert VHHH\right\rangle+A_{10}\left\vert VHHV\right\rangle+A_{11}\left\vert VHVH\right\rangle+A_{12}\left\vert VHVV\right\rangle\nonumber\\
+&A_{13}\left\vert VVHH\right\rangle+A_{14}\left\vert VVHV\right\rangle+A_{15}\left\vert VVVH\right\rangle+A_{16}\left\vert VVVV\right\rangle,
\end{align}
where the coefficients satisfy the normalized condition $\Sigma_{i=1}^{16} |A_{i}|^2=1$. To describe the processes clearly, we separate them into the following five steps.

\subsection{\textit{step 1: adding the polarization information of control single photons to the ancilla single photon. }}

Firstly, we introduce a single photon ($\left\vert H\right\rangle^a$) as ancilla and let the first single photon of the above state control the ancilla single photon through an original c-path gate, so the following state can be achieved,
\begin{align}
&\left(A_1\left\vert HHHH\right\rangle+A_2\left\vert HHHV\right\rangle+A_3\left\vert HHVH\right\rangle+A_4\left\vert HHVV\right\rangle\right.\nonumber\\
+&A_5\left\vert HVHH\right\rangle+A_6\left\vert HVHV\right\rangle+A_7\left\vert HVVH\right\rangle\nonumber\\
+&\left.A_8\left\vert HVVV\right\rangle\right)\left\vert H\right\rangle^a_1+\left(A_9\left\vert VHHH\right\rangle+A_{10}\left\vert VHHV\right\rangle\right.\nonumber\\
+&A_{11}\left\vert VHVH\right\rangle+A_{12}\left\vert VHVV\right\rangle+A_{13}\left\vert VVHH\right\rangle+A_{14}\left\vert VVHV\right\rangle\nonumber\\
+&\left.A_{15}\left\vert VVVH\right\rangle+A_{16}\left\vert VVVV\right\rangle\right)\left\vert H\right\rangle^a_2.
\end{align}
After that, the second and third single photons control the ancilla single photon in turn by the two original c-path gates respectively. The ancilla single photon will be separated into $8$ spatial modes, which depend on the polarizations of the three single photons except of the fourth single photon, as indicated in the following state,
\begin{align}
&A_1\left\vert HHHH\right\rangle\left\vert H\right\rangle^a_1+A_2\left\vert HHHV\right\rangle\left\vert H\right\rangle^a_1+A_3\left\vert HHVH\right\rangle\left\vert H\right\rangle^a_2\nonumber\\
+&A_4\left\vert HHVV\right\rangle\left\vert H\right\rangle^a_2
+A_5\left\vert HVHH\right\rangle\left\vert H\right\rangle^a_3+A_6\left\vert HVHV\right\rangle\left\vert H\right\rangle^a_3\nonumber\\
+&A_7\left\vert HVVH\right\rangle\left\vert H\right\rangle^a_4+A_8\left\vert HVVV\right\rangle\left\vert H\right\rangle^a_4+A_9\left\vert VHHH\right\rangle\left\vert H\right\rangle^a_5\nonumber\\
+&A_{10}\left\vert VHHV\right\rangle\left\vert H\right\rangle^a_{5}+A_{11}\left\vert VHVH\right\rangle\left\vert H\right\rangle^a_{6}+A_{12}\left\vert VHVV\right\rangle\left\vert H\right\rangle^a_{6}\nonumber\\
+&A_{13}\left\vert VVHH\right\rangle\left\vert H\right\rangle^a_{7}+A_{14}\left\vert VVHV\right\rangle\left\vert H\right\rangle^a_{7}\nonumber\\
+&A_{15}\left\vert VVVH\right\rangle\left\vert H\right\rangle^a_{8}+A_{16}\left\vert VVVV\right\rangle\left\vert H\right\rangle^a_{8}.
\end{align}

\subsection{\textit{step 2: the first triple-control unitary operation}}

Now, the whole polarization information of the three single photons is encoded to the spatial modes of the ancilla single photon, so that we can use the ancilla single photon to control the fourth single photon and implement the first triple-control unitary operation conveniently. We only use the spatial mode $8$ of the ancilla single photon as control signal to control the fourth single photon by the new c-path gate. According to Eq. (\ref{ncpath}), the fourth single photon will be separated into spatial mode $2$ when the spatial mode $8$ of ancilla single photon involves a single photon, otherwise, it will be separated into spatial mode $1$, i.e., the following state can be obtained,
\begin{align}
&A_1\left\vert HHHH_1\right\rangle\left\vert H\right\rangle^a_1+A_2\left\vert HHHV_1\right\rangle\left\vert H\right\rangle^a_1+A_3\left\vert HHVH_1\right\rangle\left\vert H\right\rangle^a_2\nonumber\\
+&A_4\left\vert HHVV_1\right\rangle\left\vert H\right\rangle^a_2
+A_5\left\vert HVHH_1\right\rangle\left\vert H\right\rangle^a_3+A_6\left\vert HVHV_1\right\rangle\left\vert H\right\rangle^a_3\nonumber\\
+&A_7\left\vert HVVH_1\right\rangle\left\vert H\right\rangle^a_4+A_8\left\vert HVVV_1\right\rangle\left\vert H\right\rangle^a_4+A_9\left\vert VHHH_1\right\rangle\left\vert H\right\rangle^a_5\nonumber\\
+&A_{10}\left\vert VHHV_1\right\rangle\left\vert H\right\rangle^a_{5}+A_{11}\left\vert VHVH_1\right\rangle\left\vert H\right\rangle^a_{6}+A_{12}\left\vert VHVV_1\right\rangle\left\vert H\right\rangle^a_{6}\nonumber\\
+&A_{13}\left\vert VVHH_1\right\rangle\left\vert H\right\rangle^a_{7}+A_{14}\left\vert VVHV_1\right\rangle\left\vert H\right\rangle^a_{7}\nonumber\\
+&A_{15}\left\vert VVVH_2\right\rangle\left\vert H\right\rangle^a_{8}+A_{16}\left\vert VVVV_2\right\rangle\left\vert H\right\rangle^a_{8}.
\end{align}
Only in the last two components of above equation, the fourth single photon appears on the spatial mode $2$, on which the following desired single photon unitary operation (could be arbitrary) 
\begin{align}
U_1=\left(                 
  \begin{array}{cc}   
   e^{i\delta_1} \cos\beta_1& -e^{i\psi_1}\sin\beta_1 \\  
   e^{-i\psi_1}\sin\beta_1 & e^{-i\delta_1}\cos\beta_1 \\  
  \end{array}
\right)   
\end{align}
is applied, and a merging gate is applied to the fourth single photon using the ancilla single photon as control signal. The above state will be evolved to, 
\begin{align}
&A_1\left\vert HHHH\right\rangle\left\vert H\right\rangle^a_1+A_2\left\vert HHHV\right\rangle\left\vert H\right\rangle^a_1+A_3\left\vert HHVH\right\rangle\left\vert H\right\rangle^a_2\nonumber\\
+&A_4\left\vert HHVV\right\rangle\left\vert H\right\rangle^a_2
+A_5\left\vert HVHH\right\rangle\left\vert H\right\rangle^a_3+A_6\left\vert HVHV\right\rangle\left\vert H\right\rangle^a_3\nonumber\\
+&A_7\left\vert HVVH\right\rangle\left\vert H\right\rangle^a_4+A_8\left\vert HVVV\right\rangle\left\vert H\right\rangle^a_4+A_9\left\vert VHHH\right\rangle\left\vert H\right\rangle^a_5\nonumber\\
+&A_{10}\left\vert VHHV\right\rangle\left\vert H\right\rangle^a_{5}+A_{11}\left\vert VHVH\right\rangle\left\vert H\right\rangle^a_{6}+A_{12}\left\vert VHVV\right\rangle\left\vert H\right\rangle^a_{6}\nonumber\\
+&A_{13}\left\vert VVHH\right\rangle\left\vert H\right\rangle^a_{7}+A_{14}\left\vert VVHV\right\rangle\left\vert H\right\rangle^a_{7}\nonumber\\
+&(A_{15}e^{i\delta_1} \cos\beta_1-A_{16}e^{-i\psi_1}\sin\beta_1)\left\vert VVVH\right\rangle\left\vert H\right\rangle^a_{8}\nonumber\\
+&(A_{15}e^{-i\psi_1}\sin\beta_1+A_{16}e^{-i\delta_1} \cos\beta_1)\left\vert VVVV\right\rangle\left\vert H\right\rangle^a_{8}, \label{first}
\end{align}
i.e., the first triple-control unitary operation to the fourth single photon has been completed.

\subsection{\textit{step 3: removing and adding process}}
The second triple-control unitary operation can be implemented the same as above by introducing another ancilla single photon, however, the implementation is not efficient enough. Exactly, the second triple-control unitary operation needs the polarization information of first, second and fourth single photons, while the ancilla single photon after the first triple-control unitary operation includes the polarization information of first, second and third single photons, therefore we only need to remove the polarization information of third single photon and add that of fourth single photon to the ancilla single photon and then use the ancilla single photon to control the third single photon. The removing process could be completed by a merging gate with the third single photon as control signal and the spatial modes will be merged as follows,
\begin{align}
(1,2)\rightarrow 1;  (3,4)\rightarrow 2;  (5,6)\rightarrow 3;  (7,8)\rightarrow 4,
\end{align}
The state in Eq. (\ref{first}) will be evolved back to the following state,
\begin{align}
&A_1\left\vert HHHH\right\rangle\left\vert H\right\rangle^a_1+A_2\left\vert HHHV\right\rangle\left\vert H\right\rangle^a_1+A_3\left\vert HHVH\right\rangle\left\vert H\right\rangle^a_1\nonumber\\
+&A_4\left\vert HHVV\right\rangle\left\vert H\right\rangle^a_1
+A_5\left\vert HVHH\right\rangle\left\vert H\right\rangle^a_2+A_6\left\vert HVHV\right\rangle\left\vert H\right\rangle^a_2\nonumber\\
+&A_7\left\vert HVVH\right\rangle\left\vert H\right\rangle^a_2+A_8\left\vert HVVV\right\rangle\left\vert H\right\rangle^a_2+A_9\left\vert VHHH\right\rangle\left\vert H\right\rangle^a_3\nonumber\\
+&A_{10}\left\vert VHHV\right\rangle\left\vert H\right\rangle^a_{3}+A_{11}\left\vert VHVH\right\rangle\left\vert H\right\rangle^a_{3}+A_{12}\left\vert VHVV\right\rangle\left\vert H\right\rangle^a_{3}\nonumber\\
+&A_{13}\left\vert VVHH\right\rangle\left\vert H\right\rangle^a_{4}+A_{14}\left\vert VVHV\right\rangle\left\vert H\right\rangle^a_{4}\nonumber\\
+&(A_{15}e^{i\delta_1} \cos\beta_1-A_{16}e^{-i\psi_1}\sin\beta_1)\left\vert VVVH\right\rangle\left\vert H\right\rangle^a_{4}\nonumber\\
+&(A_{15}e^{-i\psi_1}\sin\beta_1+A_{16}e^{-i\delta_1} \cos\beta_1)\left\vert VVVV\right\rangle\left\vert H\right\rangle^a_{4}. 
\end{align}
After that the adding process could be implemented by an original c-path gate with the fourth single photon as control signal and the following state can be achieved,
\begin{align}
&A_1\left\vert HHHH\right\rangle\left\vert H\right\rangle^a_1+A_2\left\vert HHHV\right\rangle\left\vert H\right\rangle^a_2+A_3\left\vert HHVH\right\rangle\left\vert H\right\rangle^a_1\nonumber\\
+&A_4\left\vert HHVV\right\rangle\left\vert H\right\rangle^a_2
+A_5\left\vert HVHH\right\rangle\left\vert H\right\rangle^a_3+A_6\left\vert HVHV\right\rangle\left\vert H\right\rangle^a_4\nonumber\\
+&A_7\left\vert HVVH\right\rangle\left\vert H\right\rangle^a_3+A_8\left\vert HVVV\right\rangle\left\vert H\right\rangle^a_4+A_9\left\vert VHHH\right\rangle\left\vert H\right\rangle^a_5\nonumber\\
+&A_{10}\left\vert VHHV\right\rangle\left\vert H\right\rangle^a_{6}+A_{11}\left\vert VHVH\right\rangle\left\vert H\right\rangle^a_{5}+A_{12}\left\vert VHVV\right\rangle\left\vert H\right\rangle^a_{6}\nonumber\\
+&A_{13}\left\vert VVHH\right\rangle\left\vert H\right\rangle^a_{7}+A_{14}\left\vert VVHV\right\rangle\left\vert H\right\rangle^a_{8}\nonumber\\
+&(A_{15}e^{i\delta_1} \cos\beta_1-A_{16}e^{-i\psi_1}\sin\beta_1)\left\vert VVVH\right\rangle\left\vert H\right\rangle^a_{7}\nonumber\\
+&(A_{15}e^{-i\psi_1}\sin\beta_1+A_{16}e^{-i\delta_1} \cos\beta_1)\left\vert VVVV\right\rangle\left\vert H\right\rangle^a_{8}. \label{pro1}
\end{align}

\subsection{\textit{step 4: the other three triple-control unitary operations}}
Obviously, the ancilla single photon in Eq. (\ref{pro1}) contains the whole polarization information of the first, second and fourth single photons, so that the second triple-control unitary operation will be applied by using the spatial mode $8$ of ancilla single photon as control signal, associated with a new c-path gate, arbitrary single photon unitary operation 
\begin{align}
U_2=\left(                 
  \begin{array}{cc}   
   e^{i\delta_2} \cos\beta_2& -e^{i\psi_2}\sin\beta_2 \\  
   e^{-i\psi_2}\sin\beta_2 & e^{-i\delta_2}\cos\beta_2 \\  
  \end{array}
\right)   
\end{align}
applied to the spatial mode $2$ of third single photon after c-path gate, and a merging gate. The following state will be obtained,
\begin{align}
&A_1\left\vert HHHH\right\rangle\left\vert H\right\rangle^a_1+A_2\left\vert HHHV\right\rangle\left\vert H\right\rangle^a_2+A_3\left\vert HHVH\right\rangle\left\vert H\right\rangle^a_1\nonumber\\
+&A_4\left\vert HHVV\right\rangle\left\vert H\right\rangle^a_2
+A_5\left\vert HVHH\right\rangle\left\vert H\right\rangle^a_3+A_6\left\vert HVHV\right\rangle\left\vert H\right\rangle^a_4\nonumber\\
+&A_7\left\vert HVVH\right\rangle\left\vert H\right\rangle^a_3+A_8\left\vert HVVV\right\rangle\left\vert H\right\rangle^a_4+A_9\left\vert VHHH\right\rangle\left\vert H\right\rangle^a_5\nonumber\\
+&A_{10}\left\vert VHHV\right\rangle\left\vert H\right\rangle^a_{6}+A_{11}\left\vert VHVH\right\rangle\left\vert H\right\rangle^a_{5}+A_{12}\left\vert VHVV\right\rangle\left\vert H\right\rangle^a_{6}\nonumber\\
+&A_{13}\left\vert VVHH\right\rangle\left\vert H\right\rangle^a_{7}+\left[A_{14}e^{i\delta_2} \cos\beta_2 \right.\nonumber\\
-&\left.(A_{15}e^{-i\psi_1}\sin\beta_1+A_{16}e^{-i\delta_1} \cos\beta_1)e^{-i\psi_2}\sin\beta_2\right]\left\vert VVHV\right\rangle\left\vert H\right\rangle^a_{8}\nonumber\\
+&(A_{15}e^{i\delta_1} \cos\beta_1-A_{16}e^{-i\psi_1}\sin\beta_1)\left\vert VVVH\right\rangle\left\vert H\right\rangle^a_{7}+\left[A_{14}e^{-i\psi_2}\sin\beta_2\right.\nonumber\\
+&\left.(A_{15}e^{-i\psi_1}\sin\beta_1+A_{16}e^{-i\delta_1} \cos\beta_1)e^{-i\delta_2} \cos\beta_2\right]\left\vert VVVV\right\rangle\left\vert H\right\rangle^a_{8}, \label{pro2}
\end{align}
which is the desired state after two triple-control unitary operations are applied.

Similarly, to implement the third triple-control unitary operation, we first merge the spatial modes as follows, \begin{align}
(1,3)\rightarrow 1;  (2,4)\rightarrow 2;  (5,7)\rightarrow 3;  (6,8)\rightarrow 4,
\end{align}
by a merge gate using the second single photon as control signal, in order to remove the polarization information of second single photon from the ancilla single photon. After that, following the process from Eq. (\ref{pro1}) to Eq. (\ref{pro2}), the third triple-control unitary operation can be completed. The final triple-control unitary operation needs the following merging operation, 
\begin{align}
(1,5)\rightarrow 1;  (3,7)\rightarrow 2;  (2,6)\rightarrow 3;  (4,8)\rightarrow 4,
\end{align}
and the similar operations discussed as above.

\subsection{\textit{step 5: removing the ancilla single photon}}
After the desired four controlled unitary operations have been completed, the ancilla single photon should be disentangled from the four-photon state, otherwise it may affect the further operations performed to the four-photon state. Since the ancilla single photon now contains the polarization information of second, third and fourth single photons, then we can use the second, third and fourth single photon as control signals to merge the ancilla single photon by the three merging gates as indicated in Fig. \ref{fig3}. After the spatial mode of ancilla single photon is resumed to be one, the ancilla single photon has been disentangled from the four-photon state.

\section{implementation of multiple ($n$-$1$)-control unitary operation} \label{sec:5}
The generalization of above approach is straightforward, and one can implement multiple ($n$) multi-control (($n$-$1$)-control) unitary operation more efficiently than the former approaches. The initial state can be described as the following form 
\begin{align}
\sum_{i=1}^{2^n}A_{i}\left\vert i\right\rangle^{(n)}=&A_1\left\vert H\cdots HH\right\rangle_{1,\cdots,n}+A_2\left\vert H\cdots HV\right\rangle_{1,\cdots,n} \nonumber\\
+&A_3\left\vert H\cdots VH\right\rangle_{1,\cdots,n}+A_4\left\vert H\cdots VV\right\rangle_{1,\cdots,n}+\cdots\label{initialn}
\end{align}
where $\sum_{i=1}^{2^n} |A_i|^2=1$. Firstly, we use the first $n-1$ single photons except of the final single photon as control signals to control an ancilla single photon $\left\vert H\right\rangle^a$ by $n-1$ original c-path gates respectively, yielding the following state,
\begin{align}
\sum_{i=1}^{2^{n-1}}A_{2i-1}\left\vert 2i-1\right\rangle^{(n)}\left\vert H\right\rangle^a_{i}+\sum_{i=1}^{2^{n-1}}A_{2i}\left\vert 2i\right\rangle^{(n)}\left\vert H\right\rangle^a_{i}.
\end{align}
To implement the first ($n$-$1$)-control unitary operation, we use the final spatial mode $2^{n-1}$ of ancilla single photon as control signal to control the last one single photon by a new c-path gate and the last one single photon will be separated into $2$ spatial modes. Applying the desired single photon unitary operation $U_1$ to the spatial mode 2 of last one single photon and merging the last one single photon by a merging gate, the desired ($n$-$1$)-control unitary operation can be implemented.

Similarly, to implement the other ($n$-$1$)-control unitary operations, the polarization information of the target single photon should be removed by merging the corresponding spatial modes through a merging gate. After that, the polarization information of the single photon, which was just as the target photon in the last control unitary operation, should be added to the ancilla single photon by an original c-path gate. A pair of new c-path and original merging gate, associated with a single photon unitary operation will complete the desired control unitary operations as well.

\section{Comparison with the former approaches}  \label{sec:6}
Now we compare the complexity of the present approach with the former approaches and the amount of required two-qubit gates are displayed in Tab. \ref{table1}. In traditional quantum computation approach, the multi-control unitary operation is usually decomposed into two-photon (e.g. CNOT) gates and single photon unitary operations. Since the single photon unitary operation can be implemented by a half wave plate and a quarter wave plate in optical system \cite{rmp}, then the amount of single photon unitary operation will not be taken into account here. It has been demonstrated that a general ($n$-$1$)-control gate requires $\mathcal{O}(n^2)$ two-qubit gates \cite{Barenco}. In this case, $\mathcal{O}(n^3)$ two-qubit gates are necessary to implement $n$ multi-control gates, i.e., the complexity is about $\mathcal{O}(n^3)$ at least.

The ($n$-$1$)-control gate can be implemented more efficiently than the decomposition approach with the original c-path and merging gate (denoted by original CPM in Tab. \ref{table1}). It has been demonstrated in Refs. \cite{lin2, lin3}, only $n$-$1$ pairs of c-path and merging gates are enough for a general ($n$-$1$)-control gate, so that the complexity for $n$ multi-control gates with the original c-path and merging gates is about $\mathcal{O}(n^2)$. 

In the present approach (denoted by new CPM in Tab. \ref{table1}), $n$-$1$ pairs of original c-path and merging gates are needed to add and remove the polarization information to the ancilla single photon firstly. For implementing each control operations, only one pair of the new c-path gate and the original merging gate, associated with one pair of original c-path and merging gates are enough. Therefore, only $3n-1$ pairs of original or new c-path and merging gates are required in the present approach, i.e., the complexity here is preserved to be $\mathcal{O}(n)$, linear increasing with the amount of multi-control unitary operations. Obviously, the present approach is more efficient than the former approaches, espcially for large scale quantum computation.

\begin{tablehere}
\caption{The amount of the necessary two-qubit operations for implementing multi-control gates with different approaches. "CNOT" denotes the approach with the decomposition to CNOT gates and single photon gates. "original CPM" denotes the approach with the original c-path and merging gate in former works and "new CPM" denotes the present approach.}\label{table1}
\vspace{-1mm}\footnotesize
\begin{center} \doublerulesep 0.1pt \tabcolsep 19.5pt
\begin{tabular}{lccccc}
\hline
CNOT\cite{Barenco} & original CPM\cite{lin2, lin3} & new CPM \\
 \hline
$\mathcal{O}(n^3)$ & $\mathcal{O}(n^2)$ & $\mathcal{O}(n)$  \\

 \hline
\end{tabular}
\end{center}

\end{tablehere}

\section{implementation of multiple ($n$-$k$)-control unitary operations} \label{sec:7}

In above, the unitary operations we discussed are ($n$-$1$)-control gates. However, a quantum circuit may include the other different control gates, e.g., ($n$-$2$)-control gates, etc. In what follows, we show our approach can be generalized to implement multiple ($n$-$k$)-control unitary operations, with $k<n$. Without loss generality, we discuss how to implementing multiple ($n$-$2$)-control gates and suppose the first $n$-$2$ single photons are the control signal of first control gate. The initial state could be described by the form in Eq. (\ref{initialn}). Similarly, we firstly introduce an single photon as ancilla and use $n$-$2$ original c-path gate to control the ancilla single photon. The following state can be achieved,
\begin{align}
&\sum_{i=1}^{2^{n-2}}A_{4i-3}\left\vert 4i-3\right\rangle^{(n)}\left\vert H\right\rangle^a_{i}+\sum_{i=1}^{2^{n-2}}A_{4i-2}\left\vert 4i-2\right\rangle^{(n)}\left\vert H\right\rangle^a_{i}\nonumber\\
+&\sum_{i=1}^{2^{n-2}}A_{4i-1}\left\vert 4i-1\right\rangle^{(n)}\left\vert H\right\rangle^a_{i}+\sum_{i=1}^{2^{n-2}}A_{4i}\left\vert 4i\right\rangle^{(n)}\left\vert H\right\rangle^a_{i}.
\end{align}
Now the first control gate can be implement by using the spatial mode $2^{n-2}$ as control signal for a new c-path gate applied to the target single photon, associated with a single photon unitary operation and an original merging gate.

Next, we begin to implement the second control gate. The polarization information of one or two single photons, which are not included in the control signal of the second control gate, should be removed from the ancilla single photon. With one or two merging gates, the removing operation will be completed. After that, the necessary polarization information of the other one or two single photons can be added to the ancilla single photon by one or two original c-path gates. Following the similar processes, the desired multiple ($n$-$k$)-control unitary operations can be implemented.   

Looks that the implementation of multiple ($n$-$k$)-cimplementontrol unitary operations is more complicated than that of multiple ($n$-$1$)-control unitary operations. More operations are required, since more than one single photon's information should be removed before applied the next control gate, if $k>1$. In other word, the complexity will be beyond the linear increasing regime, i.e., $\mathcal{O}(n)$, and gradually close to $\mathcal{O}(n^2)$ if $k$ is large. On the other hand, the complexity will be reduced with $k$ increasing, if we choose the original CPM approach to implement the operations. For a particular $k$, the complexity of the present approach will even higher than the original CPM approach. In this case, one may naturally choose the better one.  Anyway, the present approach is an important addition to the original CPM approach, and one can combine two approaches to implement a quantum circuit more efficiently.

\section{Potential applications of multiple multi-control unitary operations}  \label{sec:8}

Now we discuss the potential applications of the above multiple multi-control unitary operations. A direct application is to achieve more efficient large scale quantum computation, due to the fact that the operations can be implemented extremely efficiently with the present approach. If a quantum circuit includes the structure built by a series of multi-control unitary operations, the present approach is a better choice. The property of linear increasing with the amount of involved photon number is especially suitable for large scale quantum computation.

\begin{figure}[H]\centering
\includegraphics[scale=0.6]{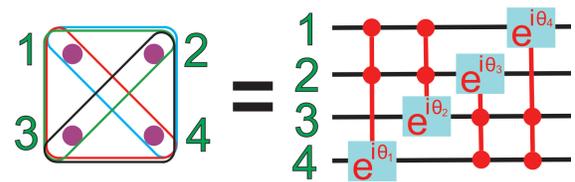}
\caption{(Color online) The 3-uniform 4-photon hypergraph state. This state can be generated by the four $2$-control phase gates in the right hand side.} 
\label{fig3}
\end{figure}

Another important application is to generate the hypergraph states \cite{hyper1,hyper2,hyper3,hyper4,hyper5}, which are regarded as the special case of locally maximally entangleable (LME) states \cite{LME}. Different to the ordinary graph states, where edges connecting two vertices, the edges in hypergraph states connect more than two vertices, so the hypergraph states are the generalized form of graph states. That does mean the structure of hypergraph state is more complicated than graph state, leading to more difficult generation of hypergraph state. Here we discuss the generation of the so-called ($n$-$k$)-uniform $n$-photon hypergraph state $\left\vert g_{n-k}\right\rangle$, which can be expressed as follows,
\begin{align}
\left\vert g_{n-k}\right\rangle=\prod_{\{i_1,\cdots,i_{n-k}\}\in E}C_{i_1,\cdots,i_{n-k}}\left\vert +\right\rangle^{\otimes n},
\end{align}
where $\{i_1,\cdots,i_{n-1}\}\in E$ means that the $k$ vertices are connected by a $k$-hyperedge, and $C_{i_1,\cdots,i_{n-k}}$ denotes the ($n$-$k$)-control phase gate. For example, in Fig. \ref{fig3} we show the $3$-uniform 4-photon hypergraph state, which can be generated with the four 2-control phase gates in the right hand side of Fig. \ref{fig3}. Obviously, the desired ($n$-$k$)-control phase gates are only the special cases of ($n$-$k$)-control unitary operations, which can be implemented efficiently by the present approach, as discussed in Sec. \ref{sec:7}. It implies that the ($n$-$k$)-uniform $n$-photon hypergraph state can be generated efficiently with our approach.

\section{Discussion and conclusions}\label{sec:9}
The core element in our approach is the XPM technique, which enables an efficient way to quantum computation. This technique recently has been used widely for quantum logic gates \cite{XPM1,XPM2,lin1,lin2,lin3,XPM3,XPM4,XPM5,lin4,lin5,logic0,logic1,logic2,logic3}, cluster or graph state generations \cite{cluster1,cluster2,cluster3,cluster4,cluster5}, entanglement concentration \cite{cen1,cen2,cen3,cen4,cen5,cen6,cen7,cen8}, entangled states generation \cite{ent1,ent2,ent3,ent4}, and others \cite{other1,other2,other3,other4,other5,other6,other7,other8,other9,other10,other11,other12}, etc. Its feasibility had been demonstrated in theory \cite{he1,he2,he3,he4,he5}, even the multi-mode effect is taken into account. More recently, the realistic experiment of an efficient XPM based on a closed-loop double-$\Lambda$ system has been reported \cite{xpme}. In addition, the similar technology has been used in the other physical system to implement a quantum computation task, such as cavity QED \cite{CQED1,CQED2,CQED3,CQED4,CQED5,CQED6,CQED7,CQED8,CQED9,CQED10,CQED11,CQED12,CQED13,CQED14,CQED15,CQED16,CQED17,CQED18,CQED19,CQED20,CQED21,CQED22,CQED23,CQED24,CQED25,CQED26,CQED27,CQED28,CQED29,CQED30,CQED31,CQED32,CQED33}, etc. Therefore the XPM technique and our present approach is feasible with the current experimental technology.

In conclusion, with the new design of c-path gate, the multiple multi-control unitary operations can be implementedimplement directly and efficiently with the combination of the original c-path, merging gates and the new c-path gates. The linear increasing with the amount of involved photon number makes this approach is suitable for large scale quantum computation.

\vspace*{2mm} \Acknowledgements{\bahao This work was supported by the National Natural Science Foundation of China (Grant No. 11574093); Natural Science Foundation of Fujian Province of China (Grant No. 2017J01004); Promotion Program for Young and Middle-aged Teacher in Science and Technology Research of Huaqiao University (Grant No. ZQN-PY113).}

\begin{appendix}
\setcounter{section}{0}
\def\thesection{Appendix}

\section{\label{sec:compint}}

\renewcommand{\theequation}{a\arabic{equation}}
\setcounter{equation}{0}

\setcounter{section}{0}
\def\thesection{Appendix A}

\begin{figure}[H]
\centering
\includegraphics[scale=0.4]{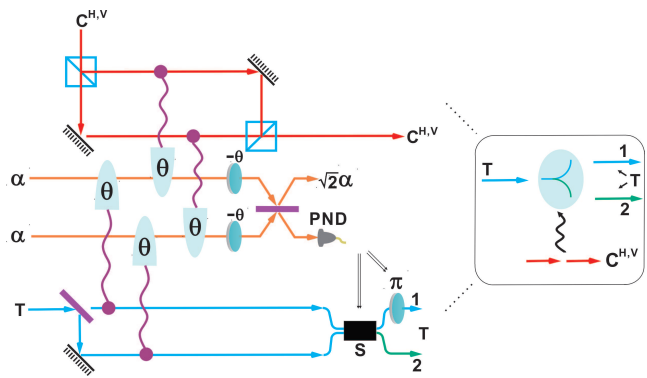}
\caption{The original c-path gate developed in Refs. \cite{lin1,lin2,lin3,lin4,lin5}. The control single photon passes through a polarization beam splitter (PBS). The $\left\vert H\right\rangle$ mode is transmitted while the $\left\vert V\right\rangle$ mode is reflected. At the same time, the target single photon passes through a 50:50 beam splitter and is separated into two spatial modes. These modes are interacted with two qubus beams as indicated in the figure. After that, two phase shifts $-\theta$ are applied to the two qubus beams respectively. The desired operation can be completed by a switch performed on the two spatial modes and a phase shift $\pi$ applied to the spatial mode $1$, controlled by the detection of the photon number-resolving detector (PND) placed on the second output modes of the BS for the two qubus beams.} 
\label{figa1}
\end{figure}

In the following, we describe the original c-path gate presented in Refs. \cite{lin1,lin2,lin3,lin4,lin5}. The initial state could be expressed as follows,
\begin{align}
\left\vert \Psi_1\right\rangle\left\vert H\right\rangle^C \left\vert \phi\right\rangle^T+\left\vert \Psi_2\right\rangle\left\vert V\right\rangle^C \left\vert \psi\right\rangle^T.
\end{align} 
Firstly, the control single photon is injected into a polarized beamsplitter (PBS), which let the $\left\vert H\right\rangle$ mode to be transmitted, while let the $\left\vert V\right\rangle$ mode to be reflected. In the same time, the target single photon is injected into a 50:50 beamsplitter (BS), and these two spatial modes associated with the two modes of control single photon interact with two qubus beams $\left\vert \alpha\right\rangle_{cs}\left\vert \alpha\right\rangle_{cs}$ as indicated in Fig. \ref{figa1}. After interaction, the following state can be achieved,
\begin{align}
&\frac{1}{\sqrt{2}}\left(\left\vert \Psi_1\right\rangle\left\vert H\right\rangle^C \left\vert \phi\right\rangle^T_1\left\vert \alpha e^{i\theta}\right\rangle_{cs}\left\vert \alpha e^{i\theta}\right\rangle_{cs}+\left\vert \Psi_1\right\rangle\left\vert H\right\rangle^C \left\vert \phi\right\rangle^T_2\left\vert \alpha\right\rangle_{cs}\left\vert \alpha e^{i2\theta}\right\rangle_{cs}\right.\nonumber\\
+&\left.\left\vert \Psi_2\right\rangle\left\vert V\right\rangle^C \left\vert \psi\right\rangle^T_1\left\vert \alpha e^{i2\theta}\right\rangle_{cs}\left\vert \alpha\right\rangle_{cs}+\left\vert \Psi_2\right\rangle\left\vert V\right\rangle^C \left\vert \psi\right\rangle^T_2\left\vert \alpha e^{i\theta}\right\rangle_{cs}\left\vert \alpha e^{i\theta}\right\rangle_{cs}\right).
\end{align} 
Two phase shift $-\theta$ are applied to the two qubus beams respectively, and let the two qubus beams to be interfered on a 50:50 BS, yielding the following state,
\begin{align}
&\frac{1}{\sqrt{2}}\left(\left\vert \Psi_1\right\rangle\left\vert H\right\rangle^C \left\vert \phi\right\rangle^T_1\left\vert \sqrt{2}\alpha \right\rangle_{cs}\left\vert 0\right\rangle_{cs}+\left\vert \Psi_1\right\rangle\left\vert H\right\rangle^C \left\vert \phi\right\rangle^T_2\left\vert \beta_{+}\right\rangle_{cs}\left\vert -\beta_{-}\right\rangle_{cs}\right.\nonumber\\
+&\left.\left\vert \Psi_2\right\rangle\left\vert V\right\rangle^C \left\vert \psi\right\rangle^T_1\left\vert \beta_{+}\right\rangle_{cs}\left\vert \beta_{-}\right\rangle_{cs}+\left\vert \Psi_2\right\rangle\left\vert V\right\rangle^C \left\vert \psi\right\rangle^T_2\left\vert \sqrt{2}\alpha \right\rangle_{cs}\left\vert 0\right\rangle_{cs}\right),
\end{align} 
where $\left\vert \beta_{-}\right\rangle_{cs}=\left\vert i\sqrt{2}\alpha \sin\theta\right\rangle_{cs}$ and $\left\vert \beta_{+}\right\rangle_{cs}=\left\vert \sqrt{2}\alpha \cos\theta\right\rangle_{cs}$. After that, we detect the second qubus beam with a photon number-resolving detector (PND). If none single photon is registered, the above state will be collapsed to
\begin{align}
\left\vert \Psi_1\right\rangle\left\vert H\right\rangle^C \left\vert \phi\right\rangle^T_1+\left\vert \Psi_2\right\rangle\left\vert V\right\rangle^C \left\vert \psi\right\rangle^T_2, \label{desa}
\end{align}
which is the desired state by the c-path gate; otherwise, the above state will be collapsed to,
\begin{align}
e^{-in\pi}\left\vert \Psi_1\right\rangle\left\vert H\right\rangle^C \left\vert \phi\right\rangle^T_2+e^{in\pi}\left\vert \Psi_2\right\rangle\left\vert V\right\rangle^C \left\vert \psi\right\rangle^T_1,
\end{align}
where $n$ is the photon number resolved by the PND. This state will be transformed to the state in Eq. (\ref{desa}) by a phase shift $\pi$ applied to the spatial mode $1$ following the classical feedforward measurement result $n$ and a switch (S) of the spatial modes $1$ and $2$. Since the induced phase shift $\theta$ is tiny, e.g., $\theta\sim 10^{-5}$, then $\sqrt{2}\alpha\cos\theta\sim\sqrt{2}\alpha$, i.e., the qubus beams could be recycled.

Now, we describe the original merging gate developed in Ref. \cite{lin5}. The initial state could be described as follows,
\begin{align}
\left\vert \Psi_1\right\rangle\left\vert H\right\rangle^C \left\vert \phi'\right\rangle^T_1+\left\vert \Psi_2\right\rangle\left\vert V\right\rangle^C \left\vert \psi'\right\rangle^T_2, \label{ime}
\end{align}
where the state of target photon could be different to the state in Eq. (\ref{desa}). Firstly, the two spatial modes of target photon are interfered on a 50:50 BS, and the second spatial mode after the BS is interacted with the second qubus beam. The above state will evolved to,
\begin{align}
&\frac{1}{\sqrt{2}}\left(\left\vert \Psi_1\right\rangle\left\vert H\right\rangle^C \left\vert \phi'\right\rangle^T_1\left\vert \alpha\right\rangle_{cs}\left\vert \alpha\right\rangle_{cs}+\left\vert \Psi_1\right\rangle\left\vert H\right\rangle^C \left\vert \phi'\right\rangle^T_2\left\vert \alpha\right\rangle_{cs}\left\vert \alpha e^{i\theta}\right\rangle_{cs}\right.\nonumber\\
+&\left.\left\vert \Psi_2\right\rangle\left\vert V\right\rangle^C \left\vert \psi'\right\rangle^T_1\left\vert \alpha\right\rangle_{cs}\left\vert \alpha\right\rangle_{cs}-\left\vert \Psi_2\right\rangle\left\vert V\right\rangle^C \left\vert \psi'\right\rangle^T_2\left\vert \alpha\right\rangle_{cs}\left\vert \alpha e^{i\theta}\right\rangle_{cs}\right).
\end{align}
Let two qubus beams to be interfered on a 50:50 BS and we can get the following state,
\begin{align}
&\frac{1}{\sqrt{2}}\left(\left\vert \Psi_1\right\rangle\left\vert H\right\rangle^C \left\vert \phi'\right\rangle^T_1\left\vert \sqrt{2}\alpha\right\rangle_{cs}\left\vert 0\right\rangle_{cs}+\left\vert \Psi_1\right\rangle\left\vert H\right\rangle^C \left\vert \phi'\right\rangle^T_2\left\vert \beta_{+}\right\rangle_{cs}\left\vert \beta_{-}\right\rangle_{cs}\right.\nonumber\\
+&\left.\left\vert \Psi_2\right\rangle\left\vert V\right\rangle^C \left\vert \psi'\right\rangle^T_1\left\vert \sqrt{2}\alpha\right\rangle_{cs}\left\vert 0\right\rangle_{cs}-\left\vert \Psi_2\right\rangle\left\vert V\right\rangle^C \left\vert \psi'\right\rangle^T_2\left\vert \beta_{+}\right\rangle_{cs}\left\vert \beta_{-}\right\rangle_{cs}\right),
\end{align}
where $\left\vert \beta_{-}\right\rangle_{cs}=\left\vert \frac{\alpha-\alpha e^{i\theta}}{\sqrt{2}}\right\rangle$ and $\left\vert \beta_{+}\right\rangle_{cs}=\left\vert \frac{\alpha+\alpha e^{i\theta}}{\sqrt{2}}\right\rangle$. Clearly, only the vacuum state $\left\vert 0\right\rangle_{cs}$ should be distinguished from the state $\left\vert \beta_{-}\right\rangle$, therefore a photon number non-resolving detector (PNND) is necessary to complete the discrimination. If none photon is registered by the PNND, the following desired state can be achieved, 
\begin{align}
\rightarrow&\left\vert \Psi_1\right\rangle\left\vert H\right\rangle^C \left\vert \phi'\right\rangle^T+\left\vert \Psi_2\right\rangle\left\vert V\right\rangle^C \left\vert \psi'\right\rangle^T; \label{desa2}
\end{align}
otherwise, by a single photon operation $\sigma_z$ applied to the control photon and a switch applied to the two spatial modes of target photon, controlled by the detection through the classical feedforward, the above state can be transformed to the desired one in Eq. (\ref{desa2}). Here the qubus beams could be recycled as well, since the induced phase shift could be tiny too.

If the initial state in Eq. (\ref{ime}) is replaced by the following state,
\begin{align}
\left\vert \Psi_1\right\rangle\left\vert 0\right\rangle^C\left\vert \Phi_1\right\rangle_1^T+\left\vert \Psi_2\right\rangle\left\vert 1\right\rangle^C\left\vert \Phi_2\right\rangle_2^T, 
\end{align} 
i.e., the state in Eq. (\ref{desire}) of main text, the following desired state 
\begin{align}
\left\vert \Psi_1\right\rangle\left\vert 0\right\rangle^C\left\vert \Phi_1\right\rangle^T+\left\vert \Psi_2\right\rangle\left\vert 1\right\rangle^C\left\vert \Phi_2\right\rangle^T, 
\end{align} 
can be achieved following the same processes from Eq. (\ref{ime}) to Eq. (\ref{desa2}). That does mean the original merging gate can work well as the inverse gate for the new c-path gate without any new design.

\begin{figure}[H]
\centering
\includegraphics[scale=0.45]{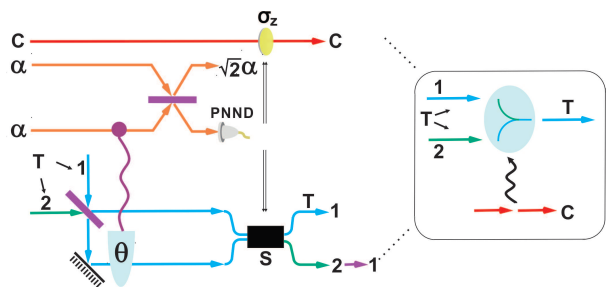}
\caption{The original merging gate developed in Ref. \cite{lin5}. The two spatial modes of target single photon are interfered on a 50:50 beam splitter. The second output mode is interacted with one of the qubus beam through the cross phase modulation process. The merging gate will be implemented after a switch performed on the two spatial modes of target single photon and a $\sigma_z$ operation applied on the control single photon, controlled by the detection of the photon number-nonresolving detector (PNND) placed on the second qubus beam.} 
\label{figa2}
\end{figure}

\end{appendix}

\end{multicols}


\begin{thebibliography}{99}
\bibitem{Nielsen}M. A. Nielsen and I. L. Chuang, Quantum Computation and Quantum Information (Cambridge University Press, Cambridge, 2000).
\bibitem{14et}T. Monz, P. Schindler, J. T. Barreiro, M. Chwalla, D. Nigg, W. A. Coish, M. Harlander, W. H\"ansel, M. Hennrich, and R. Blatt, Phys. Rev. Lett. 106, 130506 (2011).
\bibitem{8photon1}Y. F. Huang, B. H. Liu, L. Peng, Y. H. Li, L. Li, C. F. Li, and G. C. Guo, Nature commun. 2, 546 (2012).
\bibitem{8photon2}X. C. Yao, T. X. Wang, P. Xu, H. Lu, G. S. Pan, X. H. Bao, C. Z. Peng, C. Y. Lu, Y. A. Chen, and J. W. Pan, Nature Photon. 6, 225 (2012).
\bibitem{10photon}X. L. Wang, L. K. Chen, W. Li, H.-L. Huang, C. Liu, C. Chen, Y.-H. Luo, Z.-E. Su, D. Wu, Z.-D. Li, H. Lu, Y. Hu, X. Jiang, C.-Z. Peng, L. Li, N.-L. Liu, Y. A. Chen, C.-Y. Lu, and J. W. Pan,
Phys. Rev. Lett. 117, 210502 (2016).
\bibitem{KLM} E. Knill, R. Laflamme, and G. J. Milburn, Nature 409, 46 (2001).
\bibitem{CNOT1}T. B. Pittman, B. C. Jacobs, and J. D. Franson, Phys. Rev. A 64, 062311 (2001).
\bibitem{CNOT2}T. B. Pittman, M. J. Fitch, B. C. Jacobs, and J. D. Franson, Phys. Rev. A 68, 032316 (2003).
\bibitem{XPM1}S. D. Barrett, P. Kok, K. Nemoto, R. G. Beausoleil, W. J. Munro, and T. P. Spiller, Phys. Rev. A 71, 060302(R) (2005).
\bibitem{XPM2}K. Nemoto and W. J. Munro, Phys. Rev. Lett. 93, 250502 (2004).
\bibitem{lin1}Q. Lin and J. Li, Phys. Rev. A, 79, 022301 (2009).
\bibitem{lin2}Q. Lin and B. He, Phys. Rev. A 80, 042310 (2009).
\bibitem{lin3}Q. Lin, B. He, J. A. Bergou, and Y. H. Ren, Phys. Rev. A 80, 042311 (2009).
\bibitem{XPM3}W. J. Munro, K. Nemoto, T. P. Spiller, S. D. Barrett, P. Kok, and R. G. Beausoleil, J. Opt. B: Quantum Semiclass. Opt. 7, S135 (2005).
\bibitem{XPM4}W. J. Munro, K. Nemoto, and T. P. Spiller, New J. Phys. 7, 137 (2005).
\bibitem{XPM5}T. P. Spiller, K. Nemoto, S. L. Braunstein, W. J. Munro, P. van Loock, and G. J. Milburn, New J. Phys. 8, 30 (2006).
\bibitem{lin4}Q. Lin and B. He, Phys. Rev. A 82, 064303 (2010).
\bibitem{lin5}Q. Lin and B. He, Sci. Rep. 5, 12792 (2015).
\bibitem{Barenco} A. Barenco, C. H. Bennett, R. Cleve, D. P. DiVincenzo, N. Margolus, P. Shor, T. Sleator, J. A. Smolin, and H. Weinfurter, Phys. Rev. A 52, 3457 (1995).
\bibitem{hyper1} R. Qu, J. Wang, Z. Li, and Y. Bao, Phys. Rev. A 87, 022311 (2013).
\bibitem{hyper2} M. Rossi, M. Huber, D. Bru{\ss}, and C. Macchiavello, New J. Phys. 15, 113022 (2013).
\bibitem{hyper3} O. G\"uhne, M. Cuquet, F. E. S. Steinhoff, T. Moroder, M. Rossi, D. Bru{\ss}, B. Kraus, and C. Macchiavello, J. Phys. A: Math. Theor. 47, 335303 (2014).
\bibitem{hyper4} X.-Y. Chen and L. Wang, J. Phys. A: Math. Theor. 47, 415304 (2014).
\bibitem{hyper5} D. W. Lyons, D. J. Upchurch, S. N. Walck, and C. D. Yetter, J. Phys. A: Math. Theor. 48, 095301 (2015).
\bibitem{cpathe1}X. Q. Zhou, T. C. Ralph, P. Kalasuwan, M. Zhang, A. Peruzzo, B. P. Lanyon, and J. L. O'Brien, Nature Commun. 2, 413 (2011).
\bibitem{cpathe2}X. Q. Zhou, P. Kalasuwan, T. C. Ralph, and J. L. O'Brien, Nature Photon. 7, 223 (2013).
\bibitem{cpathe3}L. A. Rozema, D. H. Mahler, A. Hayat, P. S. Turner, and A. M. Steinberg, Phys. Rev. Lett. 113, 160504 (2014).
\bibitem{cpathe4}R. B. Patel, J. Ho, F. Ferreyrol, T. C. Ralph, and G. J. Pryde, Sci. Adv. 2, e1501531 (2016).
\bibitem{lin6}Q. Lin and B. He, J. Opt. B: At. Mol. Opt. Phys. 46, 055502 (2013).
\bibitem{lin7}J. L. Hu and Q. Lin, Eur. Phys. J. D, 70, 112 (2016).
\bibitem{lin8}Q. Lin, Sci. Sin.-Phys. Mech. Astron., 42, 54 (2012).
\bibitem{lin9}Q. Lin, Sci. Sin.-Phys. Mech. Astron., 44, 492 (2014).
\bibitem{lin10}Q. Lin, Sci. China-Phys. Mech. Astron., 58, 044201 (2015).
\bibitem{rmp}P. Kok, W. J. Munro, K. Nemoto, T. C. Ralph, J. P. Dowling, and G. J. Milburn, Rev. Mod. Phys. 79, 135 (2007).
\bibitem{LME} C. Kruszynska, and B. Kraus, Phys. Rev. A 79, 052304 (2009).
\bibitem{logic0}P. van Loock, W. J. Munro, K. Nemoto, T. P. Spiller, T. D. Ladd, S. L. Braunstein, and G. J. Milburn, Phys. Rev. A 78, 022303 (2008).
\bibitem{logic1}X. W. Wang, D. Y. Zhang, S. Q. Tang, L. J. Xie, Z. Y. Wang, and L. M. Kuang, Phys. Rev. A 85, 052326 (2012).
\bibitem{logic2}S. W. Lee and H. Jeong, Phys. Rev. A 87, 022326 (2013).
\bibitem{logic3}W. Zhang, P. Rui, Z. Zhang, and Q. Yang, New J. Phys. 16, 083019 (2014).
\bibitem{cluster1}S. G. R. Louis, K. Nemoto, W. J. Munro, and T. P. Spiller, Phys. Rev. A 75, 042323 (2007).
\bibitem{cluster2}S. G. R. Louis, K. Nemoto, W. J. Munro, and T. P. Spiller, New J. Phys. 9, 193 (2007).
\bibitem{cluster3}Q. Lin and B. He, Phys. Rev. A 82, 022331 (2010).
\bibitem{cluster4}Q. Lin and B. He, Phys. Rev. A 84, 062312 (2011). 
\bibitem{cluster5}Q. Lin and B. He, J. Phys. B: At. Mol. Opt. Phys. 46, 055502 (2013).
\bibitem{cen1}Y. B. Sheng and F. G. Deng, Phys. Rev. A 81, 032307 (2010).
\bibitem{cen2}Y. B. Sheng and F. G. Deng, Phys. Rev. A 81, 052323 (2010).
\bibitem{cen3}Y. B. Sheng, F. G. Deng, and G. L. Long, Phys. Rev. A 82, 032318 (2010).
\bibitem{cen4}Y. B. Sheng, L. Zhou, S. M. Zhao, and B. Y. Zheng, Phys. Rev. A 85, 012307 (2012).
\bibitem{cen5}F. G. Deng, Phys. Rev. A 85, 022311 (2012).
\bibitem{cen6}Y. B. Sheng, L. Zhou, and S. M. Zhao, Phys. Rev. A 85, 042302 (2012).
\bibitem{cen7}Y. B. Sheng and L. Zhou, Sci. Rep. 5, 7815 (2015).
\bibitem{cen8}X. H. Li and S. Ghose, Opt. Express 23, 3550 (2015).
\bibitem{ent1}X. L. Ye and Q. Lin, J. Opt. Soc. Am. B, 29, 1810 (2012).
\bibitem{ent2}Q. Lin, J. Opt. Soc. Am. B, 30, 589 (2013). 
\bibitem{ent3}Y. Zhai, Y. W. Chen, and Q. Lin, Quantum Inf. Process., 15, 761-772 (2016).
\bibitem{ent4}J. R. Hu and Q. Lin, Quantum Inf. Process., 14, 2847-2860 (2015).
\bibitem{other1}B. He, J. A. Bergou, and Y. Ren, Phys. Rev. A 76, 032301 (2007).
\bibitem{other2}B. He, Y. Ren, and J. A. Bergou, Phys. Rev. A 79, 052323 (2009).
\bibitem{other3}Q. Lin and B. He, Phys. Rev. A 80, 062312 (2009).
\bibitem{other4}Y. W. Chen and Q. Lin, Sci. China Inf. Sci. 57, 122304 (2014).
\bibitem{other5} H. F. Wang and S. Zhang, Phys. Rev. A. 79, 042336 (2009).
\bibitem{other6}Q. Guo, J. Bai, L. Y. Cheng, X. Q. Shao, H. F. Wang, and S. Zhang, Phys. Rev. A. 83, 054303 (2011). 
\bibitem{other7}H. F. Wang, A. D. Zhu, S. Zhang, and K. H. Yeon, Phys. Rev. A 87, 062337 (2013). 
\bibitem{other8} H. F. Wang, X. Q. Shao, Y. F. Zhao, S. Zhang, and K. H. Yeon, J. Opt. Soc.
Am. B. 27, 27 (2010). 
\bibitem{other9}H. F. Wang, S. Zhang, and K. H. Yeon, J. Opt. Soc. Am. B. 27, 2159 (2010). 
\bibitem{other10}H. F. Wang, S. Zhang, A. D. Zhu, X. X. Yi, and K. H. Yeon, Opt. Express 19, 25433 (2011).
\bibitem{other11}L. L. Sun, H. F. Wang, S. Zhang, and K. H. Yeon, J. Opt. Soc. Am. B. 29, 630 (2012). 
\bibitem{other12}H. F. Wang, A. D. Zhu, and S Zhang, Opt. Letters, 39, 1489 (2014).
\bibitem{he1}B. He, A. MacRae, Y. Han, A. Lvovsky, and C. Simon, Phys. Rev. A 83, 022312 (2011).
\bibitem{he2}A. Rispe, B. He, and C. Simon, Phys. Rev. Lett. 107, 043601 (2011).
\bibitem{he3}B. He, Q. Lin, and C. Simon, Phys. Rev. A 83, 053826 (2011).
\bibitem{he4}B. He and A. Scherer, Phys. Rev. A 85, 033814 (2012).
\bibitem{he5}B. He, A. V. Sharypov, J. Sheng, C. Simon, and M. Xiao, Phys. Rev. Lett. 112, 133606 (2014).
\bibitem{xpme}Z.-Y. Liu, Y.-H. Chen, Y.-C. Chen, H.-Y. Lo, P.-J. Tsai, I. A. Yu, Y.-C. Chen, and Y.-F. Chen, Phys. Rev. Lett. 117, 203601 (2016).
\bibitem{CQED1} X. Li, Y. Wu, D. Steel, D. Gammon, T. H. Stievater, D. S. Katzer, D. Park, C. Piermarocchi, and L. J. Sham, Science 301, 809 (2003).
\bibitem{CQED2} L. M. Duan and H. J. Kimble, Phys. Rev. Lett. 92, 127902 (2004).
\bibitem{CQED3}H. R. Wei and F. G. Deng, Opt. Express 21, 17671 (2013).
\bibitem{CQED4} H. R.Wei and G. L. Long, Phys. Rev. A 91, 032324 (2015).
\bibitem{CQED5} B. C. Ren, H. R. Wei, and F. G. Deng, Laser Phys. Lett. 10, 095202 (2013).
\bibitem{CQED6} B. C. Ren and F. G. Deng, Sci. Rep. 4, 4623 (2014).
\bibitem{CQED7} B. C. Ren, G. Y. Wang, and F. G. Deng, Phys. Rev. A 91, 032328 (2015).
\bibitem{CQED8} T. Li and G. L. Long, Phys. Rev. A 94, 022343 (2016).
\bibitem{CQED9} B. C. Ren and F. G. Deng, Opt. Express 25, 10863 (2017).
\bibitem{CQED10} H. R. Wei, F. G. Deng, and G.L. Long, Opt. Express 24, 18619 (2016).
\bibitem{CQED11} C. Y. Hu, W. J. Munro, and J. G. Rarity, Phys. Rev. B 78, 125318 (2008).
\bibitem{CQED12} C. Bonato, F. Haupt, S. S. R. Oemrawsingh, J. Gudat, D. Ding, M. P. van Exter, and D. Bouwmeester, Phys. Rev. Lett. 104, 160503 (2010).
\bibitem{CQED13} H. R. Wei and F. G. Deng, Sci. Rep. 4, 7551 (2014).
\bibitem{CQED14} H. R. Wei and F. G. Deng, Opt. Express 22, 593 (2014).
\bibitem{CQED15}T. van der Sar, Z. H. Wang, M. S. Blok, H. Bernien, T. H. Taminiau, D. M. Toyli, D. A. Lidar, D. D. Awschalom, R. Hanson, and V. V. Dobrovitski, Nature 484, 82 (2012).
\bibitem{CQED16} H. R. Wei, and F.G. Deng, Phys. Rev. A 88, 042323 (2013).
\bibitem{CQED17} M.J. Tao, M. Hua, Q. Ai, and F. G. Deng, Phys. Rev. A 91, 062325 (2015).
\bibitem{CQED18} X. K. Song, Q. Ai, J. Qiu, and F. G. Deng, Phys. Rev. A 93, 052324 (2016).
\bibitem{CQED19}X. K. Song, H. Zhang, Q. Ai, J. Qiu, and F. G. Deng, New J. Phys. 18, 023001 (2016).
\bibitem{CQED20} H.R. Wei, and F. G. Deng, Phys. Rev. A 87, 022305 (2013).
\bibitem{CQED21}T. J. Wang and C. Wang, Phys. Rev. A 90 , 052310 (2014).
\bibitem{CQED22}H. R. Wei and G. L. Long, Sci. Rep. 5, 12918 (2015).
\bibitem{CQED23} F. W. Strauch, Phys. Rev. A 84, 052313 (2011).
\bibitem{CQED24} C. W. Wu, M. Gao, H. Y. Li, Z. J. Deng, H. Y. Dai, P. X. Chen, and C. Z. Li, Phys. Rev. A 85, 042301 (2012).
\bibitem{CQED25}M. Hua, M. J. Tao, and F. G. Deng, Phys. Rev. A 90, 012328 (2014).
\bibitem{CQED26}M. Hua, M. J. Tao, and F. G. Deng, Sci. Rep. 5, 9274 (2015).
\bibitem{CQED27}A. Fedorov, L. Steffen, M. Baur, M. P. da Silva, and A. Wallraff, Nature 481, 170 (2012)
\bibitem{CQED28} M. Hua, M. J. Tao, F. G. Deng, and G. L. Long, Sci. Rep. 5, 14541 (2015).
\bibitem{CQED29} M. Hua, M. J. Tao, and F. G. Deng, Sci. Rep. 6, 22037 (2016).
\bibitem{CQED30}F. Schmidt-Kaler, H. H\"{a}ffner, M. Riebe, S. Gulde, G. P. T. Lancaster, T. Deuschle, C. Becher, C. F. Roos, J. Eschner, and R. Blatt, Nature 422, 408 (2003).
\bibitem{CQED31} N. A. Gershenfeld and I. L. Chuang, Science 275, 350 (1997).
\bibitem{CQED32}G. R. Feng, G. F. Xu, and G. L. Long, Phys. Rev. Lett. 110, 190501 (2013). 
\bibitem{CQED33}H. Li, Y. Liu, and G. L. Long, Sci. China-Phys. Mech. Astron., 60, 080311 (2017).
\end{thebibliography}
\end{document}